\documentclass[conference]{IEEEtran}
\IEEEoverridecommandlockouts
\usepackage{cite}
\usepackage{multirow}
\usepackage{float}
\usepackage{url}
\usepackage{amsmath,amssymb,amsfonts}
\usepackage{algorithm}
\usepackage{algorithmic}
\usepackage{algorithmic}
\usepackage{graphicx}
\usepackage{textcomp}
\usepackage{enumitem}
\usepackage{xcolor}

\def\BibTeX{{\rm B\kern-.05em{\sc i\kern-.025em b}\kern-.08em
    T\kern-.1667em\lower.7ex\hbox{E}\kern-.125emX}}
\begin{document}

\title{Dynamic Hypergraph Partitioning of Quantum Circuits with Hybrid Execution\\

}

\author{\IEEEauthorblockN{Shane Sweeney$^1$, Krishnendu Guha$^2$}\\

\IEEEauthorblockA{\textit{School of Computer Science and Information Technology} \\
\textit{University College Cork}\\
Cork, Ireland \\
sweeneyashane@gmail.com$^1$, KGuha@ucc.ie$^2$}

}

\maketitle

\begin{abstract}
Quantum algorithms offer an exponential speedup over classical algorithms for a range of computational problems. The fundamental mechanisms underlying quantum computation required the development and construction of quantum computers. These devices are referred to as NISQ (Noisy Intermediate-Scale Quantum) devices. Not only are NISQ devices extremely limited in their qubit count but they also suffer from noise during computation and this problem only gets worse as the size of the circuit increases which limits the practical use of quantum computers for modern day applications\cite{NISQ}.

This paper will focus on utilizing quantum circuit partitioning to overcome the inherent issues of NISQ devices. Partitioning a quantum circuit into smaller subcircuits has allowed for the execution of quantum circuits that are too large to fit on one quantum device. There have been many previous approaches to quantum circuit partitioning and each of these approaches differ in how they work with some focusing on hardware-aware partitioning, optimal graph-based partitioning, multi-processor architectures and many more. These approaches achieve success in their objective but they often fail to scale well which impacts cost and noise. The ultimate goal of this paper is to mitigate these issues by minimizing 3 important metrics; noise, time and cost. 

To achieve this we use dynamic partitioning for practical circuit cutting and we take advantage of the benefits of hybrid execution where classical computation will be used alongside quantum hardware. This approach has proved to be beneficial with respect to noise with classical execution enabling a 42.30\% reduction in noise and a 40\% reduction in the number of qubits required in cases where a mixture of classical and quantum computation were required.

\end{abstract}


\section{Introduction}

Quantum computing has the potential to solve problems that would have major impacts in a variety of areas. The domain of problems where quantum computers have an advantage is limited but nevertheless significant, making quantum computing a hot topic in the realm of computer science. The primary advantages that quantum computation offers are rooted in phenomena observed in quantum mechanics. Mechanisms such as superposition, entanglement, parallelism, and interference allow quantum computers to explore an exponential state space and enable efficient simulation of quantum systems\cite{book}. These mechanisms are difficult to replicate in classical computations due to infeasible memory and processing requirements. For example, while $n$ classical bits can represent one out of $2^n$ possible states at a time, $n$ qubits represent a superposition over $2^n$ states simultaneously.

\subsection{Scalability Challenges in Quantum Computing}

While there have been many algorithms designed to compute classically intractable problems, these algorithms assume that the computation is taking place on hardware that contains a large number of qubits and is fault tolerant. In reality, current quantum devices, known as NISQ (Near Intermediate-Scale Quantum) devices, are significantly limited in their qubit count and are highly sensitive to noise and decoherence between qubits. These limitations present substantial scalability challenges and are the primary reason that quantum computation has not been able to realize the potential benefits that many quantum algorithms offer. As quantum systems grow in size, maintaining coherence becomes increasingly difficult, and error rates tend to accumulate, presenting fundamental challenges to scaling quantum computations to practical problem sizes.

\subsection{Circuit Partitioning as a Solution}

One promising approach to combat the limitations of NISQ devices is to partition quantum circuits into smaller subcircuits. This approach involves taking a quantum circuit and dividing it into smaller subcircuits such that the qubit count of each subcircuit fits the constraints imposed by the hardware. Not only does this technique adhere to the physical limitations of currently available devices, but it also gives rise to the possibility of Distributed Quantum Computing (DQC)\cite{DQC}. Circuit partitioning enables larger quantum algorithms to be executed on smaller quantum processors by decomposing the original problem into manageable components that can be processed separately and later recombined.

\subsection{Existing Partitioning Approaches and Their Limitations}

Various partitioning mechanisms have been developed in recent years, each with their own strengths and weaknesses. Current approaches often rely on static partitioning schemes that fail to adapt to the unique structure of individual quantum circuits. Furthermore, many existing methods struggle with efficient inter-partition communication\cite{classical}, leading to significant overhead in quantum resources. Some approaches also face challenges in effectively determining optimal cut points within circuits efficiently, particularly for circuits with complex entanglement patterns\cite{scaling}.

\subsection{Our Proposed Approach}

This paper introduces a novel dynamic partitioning framework that adapts to circuit structure while optimizing for both quantum and classical resources. Our approach addresses the limitations of existing methods by implementing an adaptive partitioning strategy that analyzes circuit topology and qubit interaction patterns to determine efficient partition boundaries. We introduce a resource-aware offloading system that determines which subcircuits benefit most from quantum execution versus classical simulation using tensor network contractions.

The highlights of this paper can be summarized as:
\begin{enumerate}[label=(\roman*)]
    \item Dynamic partitioning that adheres partition configuration to circuit structure.
    \item Scalability in partitioning and execution. 
    \item Achieving reduced noise with noiseless classical execution.
    \item Resource-aware offloading of subcircuits for efficient use of quantum advantage.
\end{enumerate}

This paper is organized as follows. Background is discussed in Section~\ref{sec:background}. The proposed mechanism is presented in Section~\ref{sec:mechanism}. Experimentation and results are discussed in Section~\ref{sec:experiment}, and the paper concludes in Section~\ref{sec:conclusion}.

\section{Background and Motivation}
\label{sec:background}

\subsection{Circuit Partitioning Fundamentals}
Quantum circuit partitioning is the process of decomposing a large quantum circuit into smaller, self-contained subcircuits or ``partitions`` so that each can be executed independently on separate quantum processors. By mapping each qubit to a specific partition, one ensures that all gates acting on qubits within the same partition remain local, while any dependencies across partitions are removed through various circuit cutting techniques and classical recombination. This approach allows circuits exceeding the size or connectivity limits of a single device to be run on hardware with fewer qubits.
This partitioning approach uses wire cutting which involves cutting vertically through a qubits timeline\cite{cutqc}. When using wire cutting as a mechanism to partition circuits, the subcircuit upstream of the cut point must include measurements in each of the following bases at the cut point:
    
    \[
\{ I, X, Y, Z \}
\]

    Similarly, the subcircuits downstream of the cut point must be intialised in the following states at the cut point:

\[
\{\,|1\rangle,\ |0\rangle,\ |+\rangle,\ |i\rangle\,\}
\]

\begin{figure}[!htb]
  \centering
  \includegraphics[width=1\linewidth]{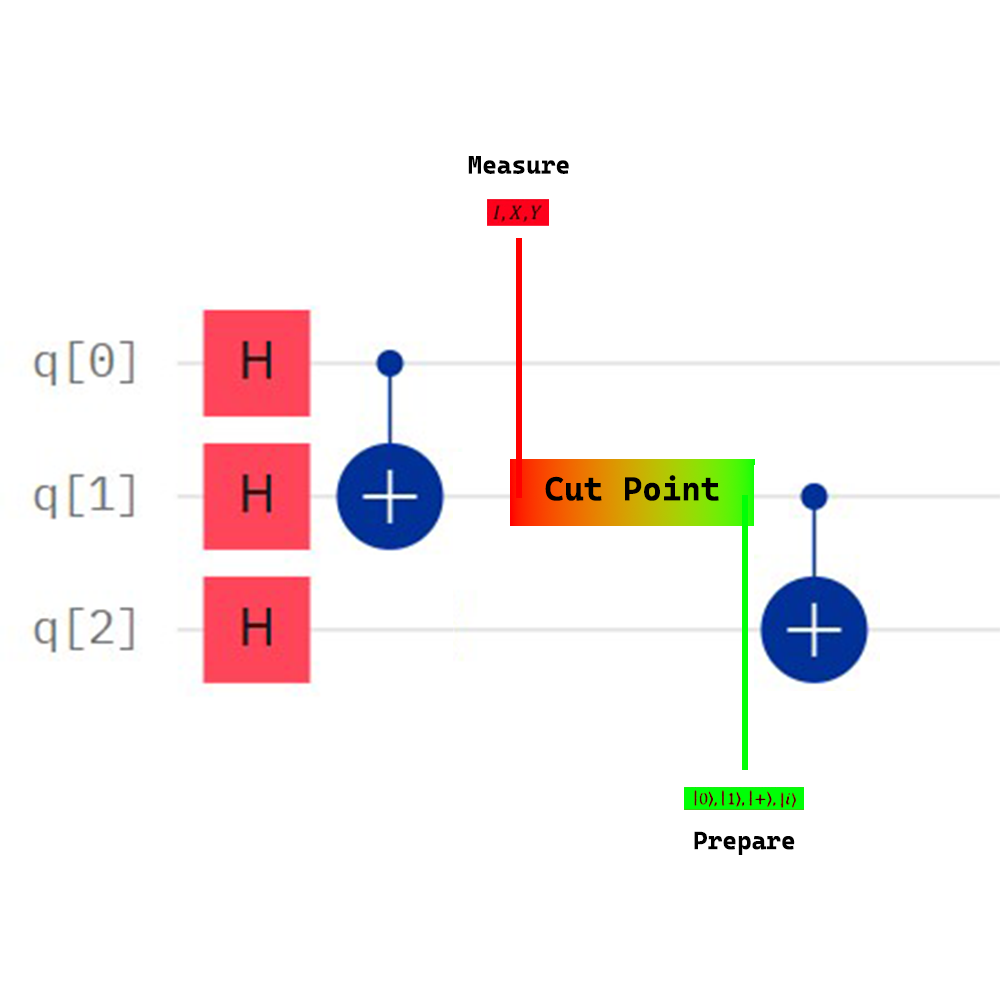}
  \caption[Wire Cut]{Wire cuts require measurements and preparations at the cut point for later reconstruction}
  \label{fig:cutting}
\end{figure}

Partitioned circuits enable distributed quantum computing by dividing the computational workload so that it can be executed across multiple quantum processing units. This distributed execution enables parallelization which is required for reducing runtime by preventing the need for serial execution which is vital in partitioning strategies that require high sampling overhead\cite{cutqc}. While there are other approaches that use interpartition communication via ebits\cite{hyper} or classical communication\cite{classical}, these approaches do not scale well to larger circuits. Enabling independent parallel execution incurs a burden during classical postprocessing\cite{cutqc} however this approach aims to minimize this burden by finding a partition configuration with minimal cut points.

\subsection{Related Works for Circuit Partitioning}
Quantum circuit partitioning approaches can be broadly categorized into three high-level domains: \textit{execution-oriented partitioning}, which adapts circuits to hardware constraints and distributed environments\cite{hwaware}; \textit{algorithmic partitioning methodologies}, which use computational and mathematical strategies to identify optimal circuit divisions\cite{optimal}; and \textit{structural decomposition}, which focuses on breaking circuits into subcomponents for simulation or error mitigation\cite{zx}. Within the algorithmic partitioning domain, cut-based techniques have become especially significant for enabling scalable and efficient quantum computation. These include strategies such as \textit{gate versus wire cutting}, where non-local gates are replaced with local operations or qubit wires are split using ancilla-assisted protocols\cite{optimal}. There are also \textit{classical communication-enhanced cuts}, which leverage intermediate measurement results and classical side information to reduce the complexity of reconstructing circuit outputs\cite{classicalcomm}. \textit{Quasiprobabilistic decomposition} enables reconstruction of global quantum states from independently executed subcircuits using probabilistic or statistical methods\cite{quasi}. Often times these approaches achieve their specific objectives but common issues that are prevalent in many of these approaches are the limitations imposed by the increases in time, cost and noise.

The approach presented in this work aims to advance cut-based partitioning by integrating dynamic hypergraph-based methods that minimize the number of cut points while preserving multi-qubit gate structures to address the exponential sampling overhead associated with traditional wire cutting\cite{scaling}. Furthermore, the framework leverages hybrid execution to combine classical tensor network simulation with quantum processing. This consolidation of advanced partitioning and hybrid execution techniques is designed to improve scalability, reduce noise, and efficiently utilize both classical and quantum resources, ultimately improving the time, cost and noise associated with quantum circuit partitioning.
~\ref{tab:previous} shows how this work combines the benefits that have been a feature in previous approaches:

\begin{table}[h!]
\caption{Table of previous approaches}
\centering
\resizebox{\columnwidth}{!}{
\begin{tabular}{|c|c|c|c|c|c|}
\hline
 & CutQC\cite{cutqc}  & Hypergraph\cite{hyper}  & GTQCP\cite{GTQCP} & Qdislib\cite{qdislib} & This Approach \\ \hline
\textbf{Dynamic number of partitions} & No & No & Yes & No & Yes \\ \hline
\textbf{Parallel Execution}     & Yes & Yes & No & Yes & Yes \\ \hline
\textbf{Hybrid execution}             & No & No & No & Yes & Yes \\ \hline
\textbf{Hypergraph Partitioning}             & No & Yes & No & No & Yes \\ \hline
\end{tabular}
}
\vspace{0.5em}
\label{tab:previous}
\end{table}

\subsection{Tensor Network Contraction}
To execute subcircuits that are deemed suitable for classical execution, an efficient framework is required to make hybrid execution feasible. Tensor networks have been found to be a very useful tool in efficiently simulating quantum circuits\cite{quantumtensor}. When using tensor networks, states are converted to tensors. These tensors then share indices at points where multi-qubit gates connect qubits. The network of tensors is then contracted to get the final output state for that subcircuit\cite{tensor}. Tensor network contraction can be improved with methods such as tensor slicing and libraries have also been developed to optimize contraction paths to improve execution time\cite{pathoptimize}. These advancements have proven that tensor networks are a powerful tool in quantum computing.

\subsection{Motivation}
The focus of this paper is to tackle scalability issues that are present in many partitioning approaches. In approaches that rely on classical post processing, an increase in cut points results in an infeasible increase in classical post processing. In approaches with inter-partition communication also fail to scale well due to latency in classical communication or the costly use of ebits. There are also other issues such as noise and cost that are inherent in all cases of quantum execution with NISQ devices. Given that this approach relies on classical post-processing\cite{tensor} to reconstruct the final result we use dynamic partitioning to minimize cut points alongside a hybrid execution approach which will improve execution time and noise.
Hybrid execution is being recognized as a promising direction for practical use of quantum advantage. Known as ``Quantum Centric Supercomputing``\cite{quantumcentric} this approach involves utilising classical supercomputers alongside quantum hardware to compute quantum algorithms.
This trend recognizes the advantage of classical resources in computing quantum circuits. The approach in this paper extends this concept further by utilising efficient classical computation to compute certain partitions. This classical execution does not override the quantum advantage as highly entangled circuits are still not feasible on classical hardware. Despite this, subcircuits with low entanglement can be executed classically where they can utilise classical benefits such as noiseless execution and scalable parallel execution. This paper aims to conceptually demonstrate that a hybrid framework could be a viable method of executing partitioned quantum circuit.

\section{Proposed Mechanism}
\label{sec:mechanism}

This section explores how this approach takes a quantum circuit as input and processes it such that it is dynamically partitioned and executed with both classical and quantum resources as outlined in ~\ref{fig:flow}.

\begin{algorithm}
\caption{Hybrid Execution Workflow}
\begin{algorithmic}[1]

\REQUIRE Quantum Circuit $QC$
\ENSURE Execution results, cost and noise metrics

\STATE Construct temporal hypergraph $G = (V, E)$ from $QC$
\FOR{$K = 2$ to $\lfloor n/2 \rfloor$}
    \STATE Partition $G$ into $K$ parts using METIS
    \STATE Refine partitions to reduce shared qubits
    \STATE Count wire cuts (cut points)
\ENDFOR
\STATE Select partitioning with minimal cut points

\STATE Generate subcircuits $S = \{s_1, s_2, \dots, s_k\}$

\FOR{each subcircuit $s_i \in S$}
    \STATE Estimate classical memory usage of $s_i$
    \IF{$s_i$ contains only classically simulable gates and low entanglement}
        \STATE Flag $s_i$ for tensor network execution
    \ELSE
        \STATE Flag $s_i$ for quantum hardware execution
    \ENDIF
\ENDFOR

\FOR{each subcircuit $s_i \in S$}
    \IF{$s_i$ flagged as classical}
        \STATE Execute using tensor network contraction
    \ELSE
        \STATE Execute on QPU or simulator
    \ENDIF
    \STATE Store execution results
\ENDFOR

\STATE Aggregate results
\STATE Compute qubit and memory savings
\STATE Estimate noise:
\[
N = S_{\text{single}} \cdot \epsilon_{\text{single}} + S_{\text{multi}} \cdot \epsilon_{\text{multi}} + Q \cdot \gamma
\]
\STATE Generate cost and noise plots

\RETURN Measurement results, noise reduction, and cost analysis

\end{algorithmic}
\end{algorithm}

\begin{figure}[!htb]
  \centering
  \includegraphics[width=1\linewidth]{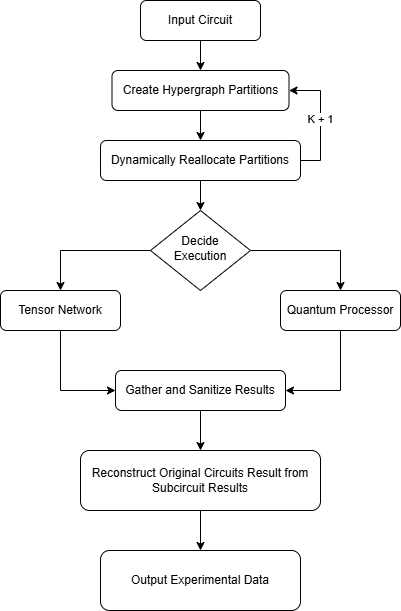}
  \caption[Wire Cut]{Flow of entire partitioning and execution framework}
  \label{fig:flow}
\end{figure}

This approach is designed to be compatible with classical post processing methods for reconstruction of the original circuits result\cite{tensor}.

\subsection{Hypergraph Construction}

A \emph{hypergraph} is a pair $G=(V,E)$ where $V$ is a set of vertices and $E\subseteq 2^V$ is a set of hyperedges, each hyperedge $e\in E$ potentially connecting more than two vertices.  In our temporal hypergraph model for quantum circuits, each node represents a qubit–time pair $v=(q_i,t_j)$, and each hyperedge $e$ represents a multi‐qubit gate acting on the set of qubits at that time.

To prepare our hypergraph for standard graph-partitioners such as METIS, we apply the \emph{clique expansion} transformation \cite{clique}. In this process, each hyperedge \(e\in E\), which may connect more than two vertices, is replaced by a complete subgraph (clique) on those same vertices. Concretely:

\[
E_x \;=\;\bigl\{(u,v)\mid \exists\,e\in E\text{ with }\{u,v\}\subseteq e\bigr\},
\]

and we assign to each new edge \((u,v)\) the average weight of all hyperedges containing both \(u\) and \(v\):

\[
w_x(u,v)
=\frac{1}{\bigl|\{\,e\in E: u,v\in e\}\bigr|}
\sum_{e:\,u,v\in e}w(e)\,.
\]

This simple averaging preserves the original multi-node connectivity strengths while producing a traditional weighted graph, enabling the use of mature graph-partitioning libraries like METIS\cite{metis}.

\paragraph{Benefits of Clique Expansion}
\begin{itemize}
  \item \textbf{Preserves multi‐qubit semantics}  Every hyperedge’s high‐order connectivity is faithfully represented by a clique among its vertices, with edge weights reflecting the original hyperedge importance\cite{hyper}.
  \item \textbf{Enables efficient tooling}  Once in $G_x$, one can apply mature graph‐partitioning libraries such as METIS or PyMetis directly, avoiding the need for specialized hypergraph partitioners.
  \item \textbf{Balances cut quality and performance}  The weighted clique expansion guides the partitioner to cut along low‐weight edges, which corresponds to minimizing the number of wire cuts in the original circuit.
\end{itemize}

\begin{figure}[!htb]
  \centering
  \includegraphics[width=1\linewidth]{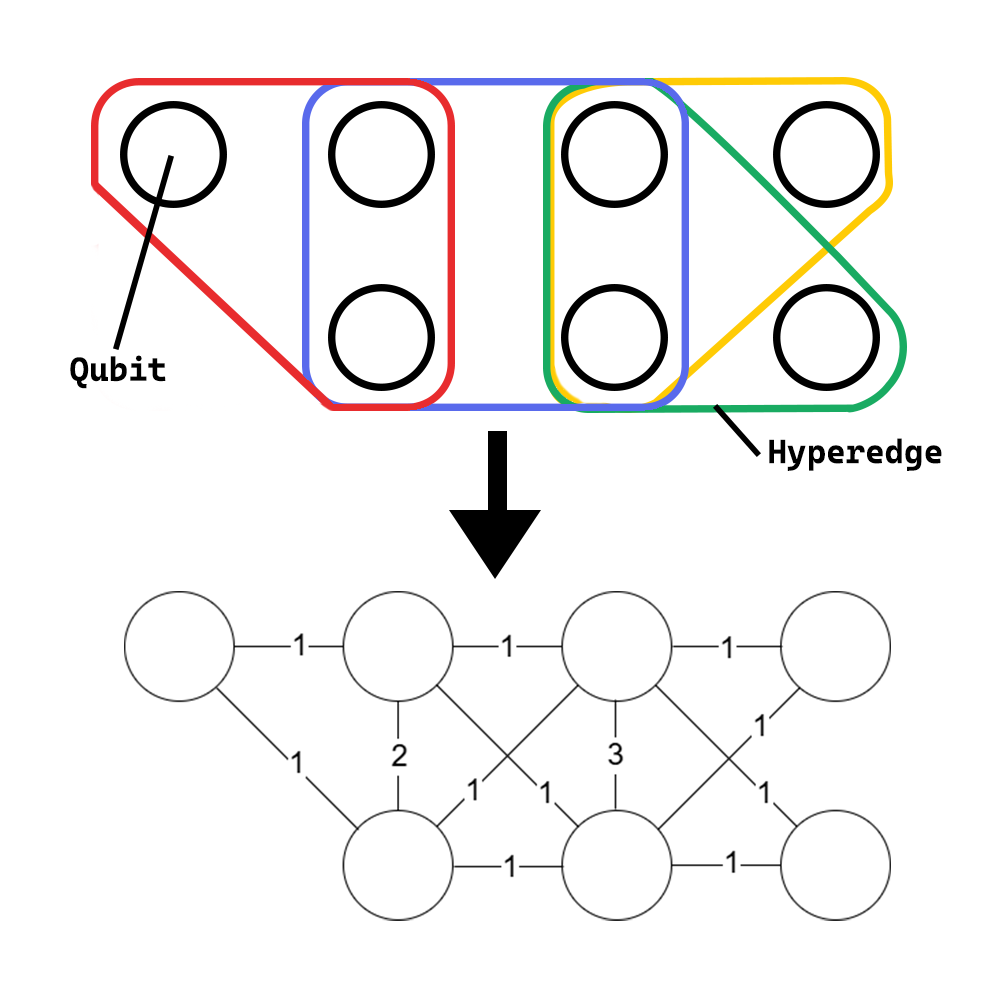}
  \caption[Wire Cut]{Diagram showing how a hypergraph is expanded to a weighted graph while respecting hypergraph semantics.}
\end{figure}

In this approach we use a temporal hypergraph. Let $G = (V, E)$ be our temporal hypergraph where:

\begin{itemize}
    \item \textbf{Nodes} $v \in V$ represent \textit{qubit-time pairs} encoded as $v = q_i\_t_j$ where:
    \begin{align*}
        q_i & : \text{Physical qubit index } (0 \leq i < n) \\
        t_j & : \text{Time step in circuit execution } (0 \leq j < D)
    \end{align*}
    This temporal node structure enables tracking of qubit state evolution.

    \item \textbf{Hyperedges} $e \in E$ represent quantum operations with:
    \[
    w(e) = \begin{cases}
    \frac{1000K}{K-1} & \text{per pairwise connection for \(K\)-qubit gates} \\
    10 & \text{for temporal continuity edges}
    \end{cases}
    \]
    
\end{itemize}

Hyperedges are expanded into complete subgraphs for partitioning while preserving hypergraph semantics. Partitioning uses graph-based tools on the clique-expanded hypergraph. The weighted hyperedge formulation aims to ensure that only wires are cut. Preventing gate cuts allows for parallel execution of each subcircuit independently. 
Once we have the graph, we use METIS partitioning to produce the subcircuits\cite{metis}. METIS takes the weighted graph derived from the temporal hypergraph and divides it into $K$ balanced partitions while minimizing the total weight of edges (cuts) between them. 

\begin{figure}[!htb]
  \centering
  \includegraphics[width=1\linewidth]{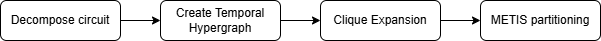}
  \caption[Wire Cut]{Flow of hypergraph-based partitionoing process}
\end{figure}

\subsection{Sampling Overhead and Mitigation}

While quantum circuit cutting has emerged as a promising strategy to enable the execution of large circuits on limited hardware by dividing a quantum circuit into smaller, hardware-compatible subcircuits, a significant challenge associated with this approach is the \textit{sampling overhead} introduced by the need to reconstruct the output of the full circuit from the independently executed subcircuits\cite{cutqc}.

When a circuit is partitioned, any qubit wire that is cut introduces a need for resolution of state at the boundary. Specifically, the upstream subcircuit must be measured in a complete set of bases, $\{I, X, Y, Z\}$ and the downstream subcircuit must be initialized accordingly using $\{|1\rangle,\ |0\rangle,\ |+\rangle,\ |i\rangle\}$ as mentioned previously. The final output is reconstructed by combining the results across all basis combinations using classical post-processing techniques.

This process leads to a \textit{sampling explosion}, where the total number of required executions scales exponentially with the number of cut points. If $C$ represents the number of wires cut, and each cut requires $B$ basis configurations where typically $B = 4$, then the total number of executions scales as:
\[
\text{Sampling Overhead} = O(B^C)
\]
Even for moderately sized circuits, a small increase in $C$ can lead to infeasible execution and reconstruction costs\cite{scaling}, especially on NISQ-era devices with limited qubit counts and access constraints. As such, reducing the number of cut points is essential for practical implementation of circuit cutting strategies.

To address this, our approach incorporates a \textit{dynamic partitioning mechanism} that explicitly targets the minimization of sampling overhead. This is achieved through several key components:

\begin{itemize}
    \item \textbf{Adaptive Partitioning Range:} Rather than fixing the number of partitions $K$ beforehand, we iteratively test values of $K$ ranging from $2$ to $\lfloor n/2 \rfloor$, where $n$ is the number of qubits in the circuit. For each candidate $K$, a METIS-based partitioning is performed on a hypergraph representation of the circuit, and the resulting number of cut points is evaluated. The configuration with the lowest $C$ is selected for execution.

    \item \textbf{Gate-Aware Hypergraph Construction:} The hypergraph model assigns large weights to multi-qubit gates to discourage gate fragmentation during partitioning. This ensures that logically connected gates remain in the same subcircuit and reduces unnecessary entanglement across partitions.

    \item \textbf{Refinement Phase:} After initial partitioning, a refinement step analyzes shared qubits and reallocates single-qubit gates to the dominant subcircuit if all multi-qubit interactions for that qubit reside in one partition. This reduces the number of qubits that appear in multiple subcircuits and further decreases $C$.

\end{itemize}

The combined effect of these strategies has demonstrated that some circuits benefit from a dynamic number of partitions to reduce cut points. This leads to fewer required executions and more efficient reconstruction, making the approach scalable and suitable for hybrid quantum-classical architectures.

\subsection{Dynamic Partitioning}
This partitioning approach combines Metis graph splitting with real-time resource monitoring. Metis partitions hypergraphs by coarsening them into smaller graphs, computing an initial partition on the smallest graph and then refining the partition while projecting it back to the original graph to minimize edge cuts and balance vertex weights. The result is $K$ partitions that provide balanced partitions that adhere to the weighting scheme imposed during hypergraph construction.

Unlike previous approaches, this approach tests multiple values of $K$. Given a circuit, the minimum value for $K$ is 2 and the maximum value for $K$ is $numqubits/2$. All values of $K$ in this range are tested until an optimal partition configuration is found.

For this project, the optimal partition is defined as the partition that produces the minimum number of cut points i.e. points at which qubits are shared across partitions due to a wire cut. This metric was chosen for optimality to overcome the sampling and classical reconstruction overhead that is associated with cutting\cite{scaling}.

\subsection{Refinement}
The reallocation phase refines the partitions by minimising the number of shared qubits where possible. This is done by analysing each qubit that is shared across subcircuits. If there is a case where a shared qubit is found to have all of its multi qubit gates in one subcircuit, all of the single qubit gates associated with that qubit in other subcircuits are reallocated into the partition with the multi-qubit gates. If there is chance for reallocation then this results in partitions with less shared qubits.

\begin{algorithm}[H]
\caption{Reallocation Phase for Minimizing Shared Qubits}
\begin{algorithmic}[1]
\REQUIRE Set of subcircuits $\mathcal{S}$, each with assigned qubits and gates
\ENSURE Refined partitions with minimized shared qubits

\FOR{each qubit $q$ shared across multiple subcircuits}
    \STATE Identify the set of subcircuits $\mathcal{S}_q$ containing $q$
    \FOR{each subcircuit $s \in \mathcal{S}_q$}
        \STATE Count the number of multi-qubit gates involving $q$ in $s$
    \ENDFOR
    \IF{all multi-qubit gates involving $q$ are in a single subcircuit $s^*$}
        \FOR{each subcircuit $s \in \mathcal{S}_q$, $s \neq s^*$}
            \STATE Move all single-qubit gates on $q$ from $s$ to $s^*$
        \ENDFOR
        \STATE Update partitions: $q$ is now only in $s^*$
    \ENDIF
\ENDFOR

\RETURN Refined set of subcircuits with reduced shared qubits
\end{algorithmic}
\end{algorithm}

\subsection{Deciding Execution}
The execution mode decision process employs a multi-criteria approach to optimize fidelity and resource utilization. The approach first determines the resources available on the local machine and then it uses these device specifications to determine the threshold for classical execution using the following criteria:

\begin{itemize}

    \item \textbf{Qubit Count Thresholding}:
    The qubit count in a subcircuit can severely limit the efficacy of classical execution due to the inherent exponential memory costs that are required to compute the state of a quantum system. For the purpose of this paper we will consider the cost of the full state when calculating memory requirements. The number of qubits that can be executed is then determined by the amount of memory available to the user.

    \item \textbf{Quantum Friendly Gates}:
    If there are non-clifford gates, the subcircuit is immediately flagged for quantum execution as these gates are not suitable for computation with tensor networks because they introduce operations that cannot be simulated easily.
    \item \textbf{Entanglement}:
    If the number of multi-qubit gates in a subcircuit is too high, the subcircuit is flagged for quantum execution as tensor networks are inefficient for highly entangled systems due to the increased size of the bond dimensions.

\end{itemize}

\begin{figure}[!htb]
  \centering
  \includegraphics[width=1\linewidth]{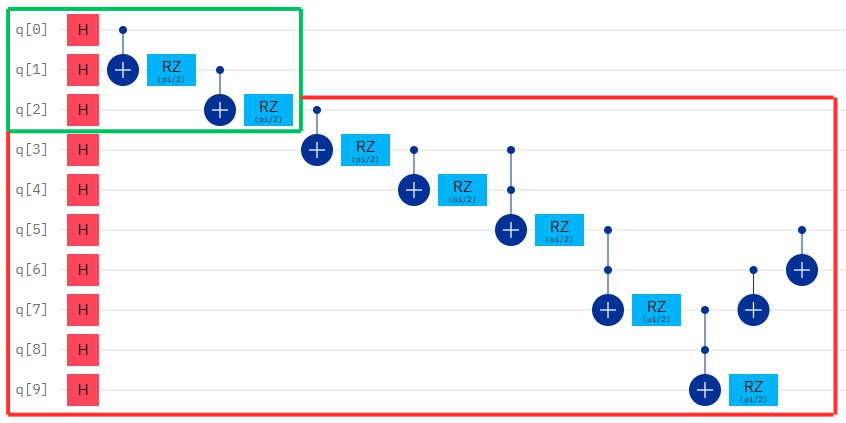}
  \caption[Wire Cut]{A hypothetical circuit for illustrative purposes highlighting how a subcircuit may qualify for quantum execution.}
  \label{fig:decision}
\end{figure}

This approach improves upon static threshold methods by dynamically adapting to the classical resources available to the user. ~\ref{fig:decision} shows a hypothetical circuit partitioned into 2 subcircuits. In this case, the threshold for quantum execution can be defined as $qubits>5$ or $multiqubit gates>5$ and so the green subcircuit would be executed classically and the red subcircuit would be executed with quantum hardware due to qubit count and multi-qubit gate count. The threshold for number of qubits and number of multi-qubit gates is defined by the amount of memory and processing power available to the user.

\subsection{Execution}

\subsubsection{Classical Execution}

The classical execution path uses tensor network contraction. The process begins by using the quimb tensor library to take the subcircuit that has been flagged for quantum execution and convert it to a tensor network.
This approach allows classical systems to efficiently simulate quantum circuits that would otherwise require prohibitive computational resources. This enables execution of quantum circuits up to the boundaries of quantum advantage. For larger subcircuits, it is possible to employ the cotengra library\cite{cotengra} to optimize contraction paths which is crucial in maintaining computational efficiency for classical execution. 
Tensor network contraction can also leverage GPU acceleration and it is also possible to alter precision strategies if there is need to reduce processing requirements at the expense of accuracy. This methodology provides a scalable framework to execute quantum circuits in an efficient manner with classical resources. 

\subsubsection{Quantum Execution}

Alternatively, subcircuits are flagged for quantum execution. In quantum execution, the circuit is first transpiled and optimised using qiskits built in optimisation and a job is sent to the IBM qiskit platform. This process takes time due to the queue time however this partitioning approach is best analysed under the assumption that there is quantum hardware readily available to the user. This ``quantum acceleration`` framework will likely be realized within the next few years as the development of quantum hardware accelerates.
Once the execution is complete, results must be gathered and sanitized to ensure type consistency for reconstruction of the final state. This approach aims to minimize cut points under the assumption that reconstruction will be done with efficient tensor network contraction\cite{tensor} however this ignores the possibility other reconstruction methods such as approximate reconstruction which aims to tackle the issue of classical overhead\cite{HMC}.

\section{Experimentation and Result Analysis}
\label{sec:experiment}

\subsection{Experimental Strategy}
This investigation used IBM's Qiskit platform to access quantum processing units. All quantum executions used 1000 shots. For classical execution, 1 laptop with Processor	Intel(R) Core(TM) i7-3667U CPU @ 2.00GHz, 2001 Mhz, 2 Core(s), 4 Logical Processor(s) and 8 GB RAM was used. For classical execution, we are using the Quimb tensor library to ensure later compatibility with the Cotengra library for optimized contraction paths. In these experiments 4 ciruits are used for testing:

\begin{itemize}
    \item \textbf{BV\cite{bv}}: The Bernstein–Vazirani oracle circuit that learns an $n$-bit secret string with a single query. 
    \item \textbf{GHZ\cite{ghz}}: An $n$-qubit Greenberger–Horne–Zeilinger state preparation circuit using one Hadamard and a cascade of CNOT gates.
    \item \textbf{Random\cite{random}}: A depth-$d$ circuit on $n$ qubits composed of randomly chosen single- and two-qubit gates provided via qiskit.
    \item \textbf{QFT\cite{qft}}: The $n$-qubit quantum Fourier transform circuit built from Hadamard and controlled-phase rotations.
\end{itemize}

\subsection{Result Analysis}
\subsubsection{Execution Time}

Table~\ref{tab:circuit_depth} demonstrates execution times for circuits of different size in terms of the number of qubits and depth. As expected, the results show an increase in execution time as the size of the circuit increases. This increase is not a major issue however it is important to consider that this approach assumes completely parallel execution of subcircuits. Serial execution of subcircuits would incur a notable penalty in terms of execution time so it is important that execution is distributed as much as possible.

\begin{figure}[!htb]
  \centering
  \includegraphics[width=1\linewidth]{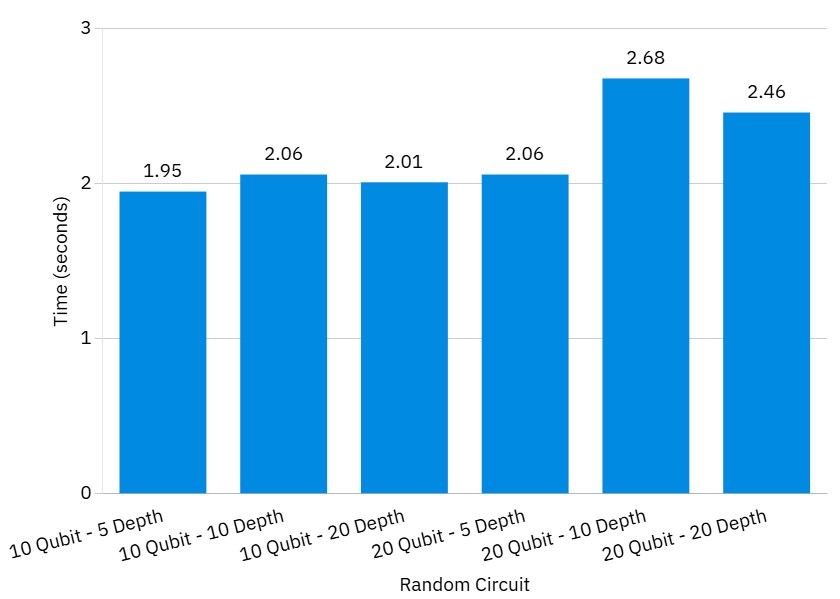}
  \caption[Wire Cut]{Maximum execution time of a subcircuit on quantum hardware via Qiskit platform}
  \label{fig:flow}
\end{figure}

\begin{table}[htbp]
\caption{Quantum Circuit Execution Times (Seconds)}
\begin{center}
\begin{tabular}{|c|c|c|c|c|}
\hline
\textbf{Circuit} & \textbf{Qubits} & \textbf{5} & \textbf{10} & \textbf{20} \\
\hline
\multirow{2}{*}{Random}
    & 10 & 1.95 & 2.06 & 2.01 \\
    & 20 & 2.06 & 2.68 & 2.46 \\
\hline
\end{tabular}
\label{tab:circuit_depth}
\end{center}
\end{table}

Table~\ref{tab:tensor_contraction} displays the maximum time observed when executing a subcircuit using tensor network contraction. Maximum execution time is important here as this process can be parallelized and distributed across compute nodes. While the tensor network execution times do exhibit the efficiency of this approach, it is important to consider that the quantum execution times take considerably longer than their classical counterpart. This suggests that more work could be allocated to the classical pathway to reduce idle time and produce a more balanced distribution of work.

\begin{figure}[!htb]
  \centering
  \includegraphics[width=1\linewidth]{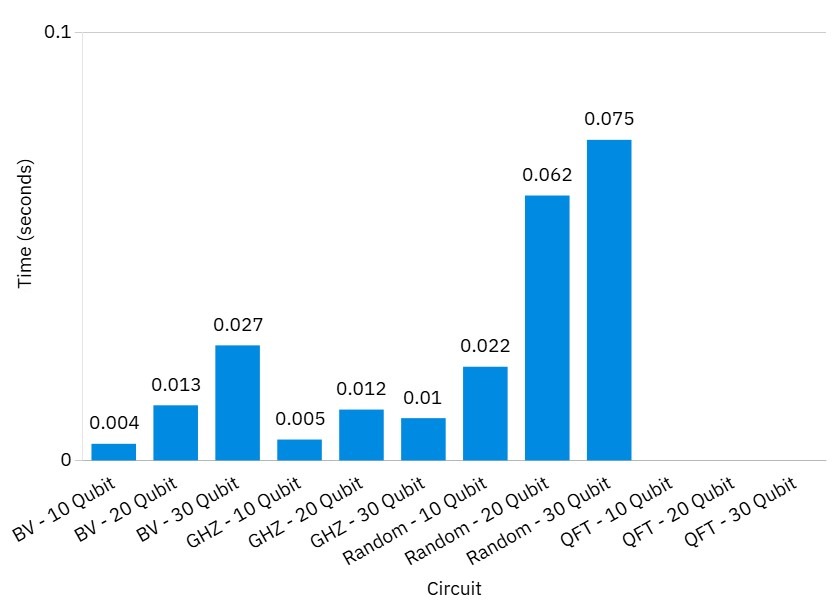}
  \caption[Wire Cut]{Maximum time required for tensor network contraction of a subcircuit}
  \label{fig:flow}
\end{figure}

\begin{table}[htbp]
\caption{Tensor Network Contraction Execution Times (Seconds)}
\begin{center}
\begin{tabular}{|c|c|c|c|}
\hline
\textbf{Circuit} & \textbf{10} & \textbf{20} & \textbf{30} \\
\hline
BV     & 0.0040 & 0.0130 & 0.0270 \\
GHZ    & 0.0050 & 0.0120 & 0.0100 \\
Random & 0.0222 & 0.0620 & 0.0750 \\
QFT    & --     & --     & -- \\
\hline
\multicolumn{4}{l}{$^{\mathrm{*}}$No times for cases of pure quantum execution}
\end{tabular}
\label{tab:tensor_contraction}
\end{center}
\end{table}

The tensor network execution results support the fact that tensor networks are an efficient mechanism when executing quantum circuits. When combined with quantum execution it becomes evident that quantum execution time is the defining factor when considering total execution time. As previously mentioned, the results are produced with a focus on parallel execution. It is important to note that while parallel execution on classical hardware is not a concern, parallel quantum execution will require several quantum processing units.

\subsection{Cut Time}
Table~\ref{tab:cut_time} reports the time required to produce the partitions. Total time represents the total time required to execute the partitioner while the Max Time shows the maximum time observed when attempting a value for $K$ which is important when considering a parallel implementation.
For each circuit given as input, the maximum partition size is determined by $n/2$ where $n$ is the number of qubits. When obtaining results, a maximum limit of 8 was placed on $K$ when gathering the following results as the dynamic partitioning often found optimal values for $K$ to be less than 5 but this maximum limit can be ignored if more constraints are later added.

\begin{figure}[!htb]
  \centering
  \includegraphics[width=1\linewidth]{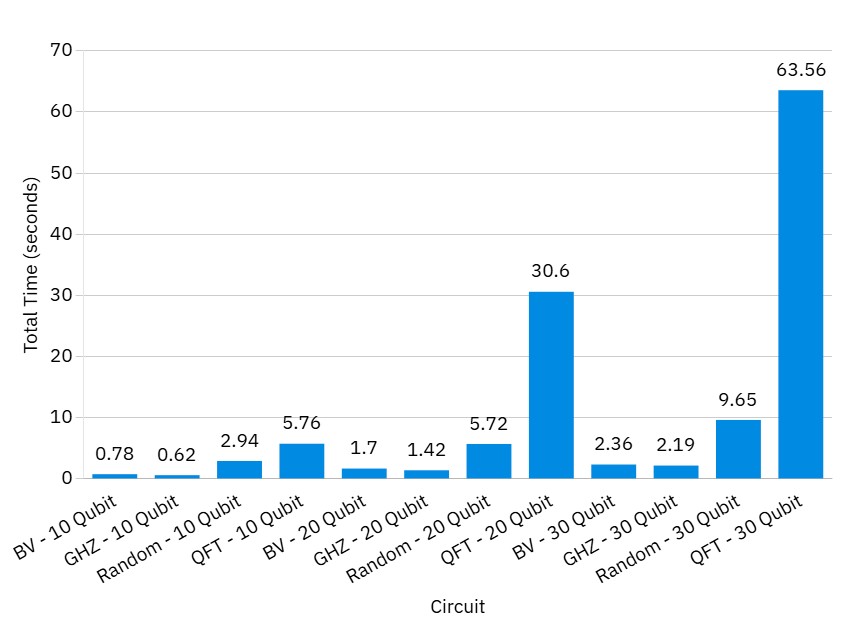}
  \caption[Wire Cut]{Total cut time in seconds}
  \label{fig:flow}
\end{figure}

\begin{figure}[!htb]
  \centering
  \includegraphics[width=1\linewidth]{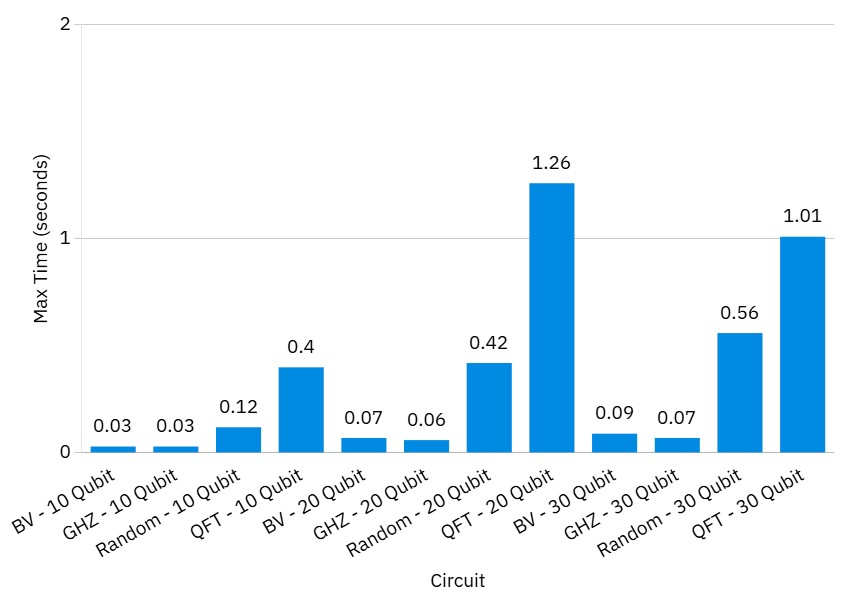}
  \caption[Wire Cut]{Maximum observed cut time for any value $K$ tried}
  \label{fig:flow}
\end{figure}

\begin{table}[H]
\caption{Cut Time Results}
\begin{center}
\begin{tabular}{|c|c|c|c|}
\hline
\textbf{Circuit} & \textbf{Size} & \textbf{Total Time} & \textbf{Max Time} \\
\hline
\multirow{3}{*}{BV}
    & 10 & 0.78 & 0.03 \\
    & 20 & 1.70 & 0.07 \\
    & 30 & 2.36 & 0.09 \\
\hline
\multirow{3}{*}{GHZ}
    & 10 & 0.62 & 0.03 \\
    & 20 & 1.42 & 0.06 \\
    & 30 & 2.19 & 0.07 \\
\hline
\multirow{3}{*}{Random}
    & 10 & 2.94 & 0.12 \\
    & 20 & 5.72 & 0.42 \\
    & 30 & 9.65 & 0.56 \\
\hline
\multirow{3}{*}{QFT}
    & 10 & 5.76 & 0.40 \\
    & 20 & 30.60 & 1.26 \\
    & 30 & 63.56 & 1.01 \\
\hline
\end{tabular}
\label{tab:cut_time}
\end{center}
\end{table}

BV and GHZ circuits exhibit low total and max cut times due to their simplistic structure. Random circuits show higher times with a time of 9.65 seconds for 30 qubit random circuit compared to 2.19 seconds for the 30 qubit GHZ circuit. QFT has the highest cut times with 63.56 seconds total for the 30 qubit configuration which is expected due to its complex, recursive structure.
It is clear from the results that the the type of circuit is the largest influence of cut times which shows degrading performance as circuits get more complex. Despite this, it is important to remember that this process can be parallelized. Parallelizing this strategy involves distributing the evaluation for each value of $K$ which would reduce the total partitioning time. 

\subsection{Noise}
Table~\ref{tab:noise} highlights the reduction in noise observed as a result of noiseless executions of subcircuits deemed suitable for classical execution. A noise score is produced as a combination of gate error contributions and crosstalk effects, calculated as:

\begin{equation}
N = S_{\text{single}} \cdot \epsilon_{\text{single}} + S_{\text{multi}} \cdot \epsilon_{\text{multi}} + Q \cdot \gamma
\end{equation}

where:
\begin{align*}
    S_{\text{single}} & : \text{Number of single-qubit gates} \\
    S_{\text{multi}}  & : \text{Number of multi-qubit gates} \\
    Q                 & : \text{Number of qubits involved} \\
    \epsilon_{\text{single}} & = 0.001, \quad \epsilon_{\text{multi}} = 0.01 : \text{Per-gate error rates} \\
    \gamma            & = 0.05 : \text{Crosstalk factor per qubit}
\end{align*}

This formula captures both gate-level errors and qubit-crosstalk effects and the coefficients were chosen carefully to represent the error rates expected in current devices.\cite{error} The purpose of this is to demonstrate the reduction in noise that is observed when executing subcircuits classically. It must be noted that classical execution can be configured to be more efficient at the expense of accuracy but for the purpose of this analysis it will be assumed that the classical computation is noiseless.

\begin{figure}[!htb]
  \centering
  \includegraphics[width=1\linewidth]{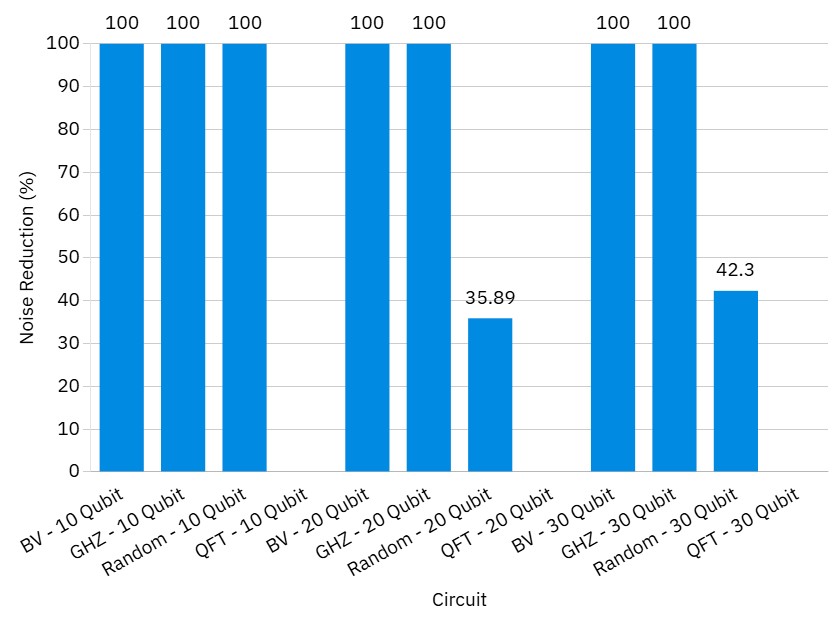}
  \caption[Wire Cut]{Noise score reductions as a result of classical execution}
  \label{fig:flow}
\end{figure}

\begin{table}[htbp]
\caption{Noise Results}
\begin{center}
\begin{tabular}{|c|c|c|c|c|}
\hline
\textbf{Circuit} & \textbf{Size} & \textbf{Original} & \textbf{Classical} & \textbf{Saved (\%)} \\
\hline
\multirow{3}{*}{BV}
    & 10 & 0.58 & 0.58 & 100 \\
    & 20 & 1.14 & 1.14 & 100 \\
    & 30 & 1.71 & 1.71 & 100 \\
\hline
\multirow{3}{*}{GHZ}
    & 10 & 0.59 & 0.59 & 100 \\
    & 20 & 1.19 & 1.19 & 100 \\
    & 30 & 1.79 & 1.79 & 100 \\
\hline
\multirow{3}{*}{Random}
    & 10 & 1.07 & 1.07 & 100 \\
    & 20 & 1.93 & 0.69 & 35.89 \\
    & 30 & 2.53 & 1.07 & 42.30 \\
\hline
\multirow{3}{*}{QFT}
    & 10 & 1.70 & 0    & 0 \\
    & 20 & 5.69 & 0    & 0 \\
    & 30 & 11.98 & 0   & 0 \\
\hline
\end{tabular}
\label{tab:noise}
\end{center}
\end{table}

For BV and GHZ circuits, classical execution achieves 100\% noise savings for all sizes of circuit. In this case each subcircuit is flagged for classical execution which results in a noiseless execution. Random circuits show partial savings in the 20 and 30 qubit configurations indicating that some subcircuits were flagged for classical execution and others for quantum execution. QFT circuits show 0\% savings as their entanglement structure resists classical simulation, requiring full quantum execution.
The results show that hybrid execution can be an effective means at reducing noise however the efficacy of this approach is limited both by the availability of powerful classical resources and the entanglement structure of the circuit being partitioned.

\subsection{Cost}

Table~\ref{tab:cost} includes results that demonstrate the reduced resources required when utilising a hybrid execution approach. The table shows the number of qubits required for executing the original circuit alongside the maximum number of qubits required to execute the largest quantum subcircuit.
The table also shows the memory requirements for executing the original circuit alongside the maximum amount of memory required to execute the largest classical subcircuit with the memory requirement calculated as:

\begin{equation}
M = B \times 2^{n}
\end{equation}

where $M$ is the total memory (in bytes), $B$ = 16 is the bytes allocated per complex number (assuming double-precision complex values\cite{complex}), and $n$ is the number of qubits. This formula represents the memory needed to store the full quantum state vector during classical simulation, reflecting the exponential scaling of memory resources with respect to qubit count. This calculation does not reflect the actual memory requirements for executing a given circuit however it does consider the number of qubits in a subcircuit which is the primary factor in estimating memory requirements. The results from this calculation therefore suffice in highlighting the differences in memory requirements between full circuit and subcircuit execution.

\begin{figure}[!htb]
  \centering
  \includegraphics[width=1\linewidth]{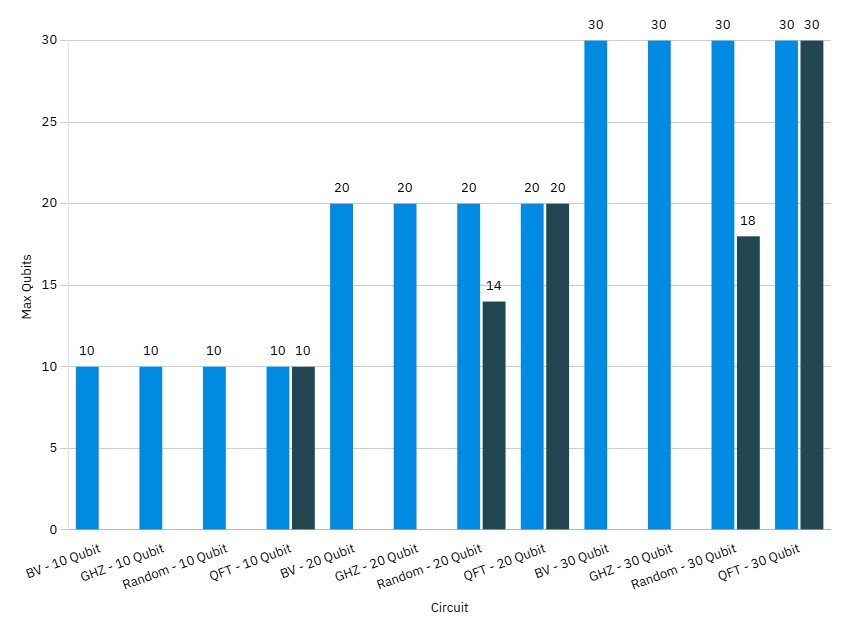}
  \caption[Wire Cut]{The maximum number of qubits required as a result of classical subcircuit execution}
  \label{fig:flow}
\end{figure}

\begin{figure}[!htb]
  \centering
  \includegraphics[width=1\linewidth]{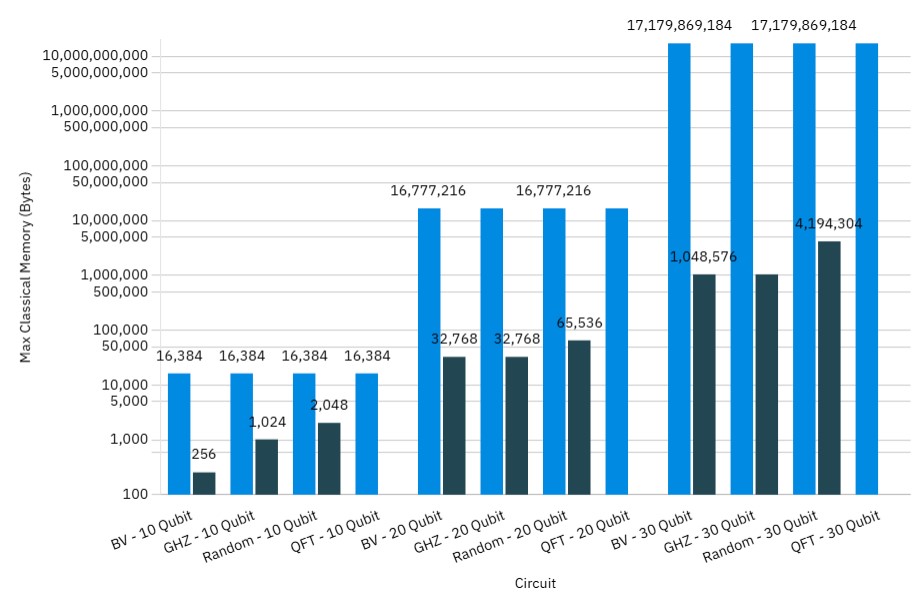}
  \caption[Wire Cut]{The maximum amount of memory required as a result of offloading quantum friendly subcircuits}
  \label{fig:flow}
\end{figure}

\begin{table}[htbp]
\caption{Cost Results}
\begin{center}
\resizebox{\columnwidth}{!}{
\begin{tabular}{|c|c|c|c|c|c|}
\hline
\textbf{Circuit} & \textbf{Size} & \textbf{Classical Total} & \textbf{Classical Max} & \textbf{Qubit Total} & \textbf{Qubit Max} \\
\hline
\multirow{3}{*}{BV}
    & 10 & 16384 & 256 & 10 & 0 \\
    & 20 & 16777216 & 32768 & 20 & 0 \\
    & 30 & 17179869184 & 1048576 & 30 & 0 \\
\hline
\multirow{3}{*}{GHZ}
    & 10 & 16384 & 1024 & 10 & 0 \\
    & 20 & 16777216 & 32768 & 20 & 0 \\
    & 30 & 17179869184 & 1048576 & 30 & 0 \\
\hline
\multirow{3}{*}{Random}
    & 10 & 16384 & 2048 & 10 & 0 \\
    & 20 & 16777216 & 65536 & 20 & 14 \\
    & 30 & 17179869184 & 4194304 & 30 & 18 \\
\hline
\multirow{3}{*}{QFT}
    & 10 & 16384 & 0 & 10 & 10 \\
    & 20 & 16777216 & 0 & 20 & 20 \\
    & 30 & 17179869184 & 0 & 30 & 30 \\
\hline
\end{tabular}
}
\label{tab:cost}
\end{center}
\end{table}

With all BV and GHZ subcircuits being executed classically, there is no need for quantum resources. Even when these subcircuits are being executed classically, there is a significant decrease in the amount of memory required to execute a subcircuit classically when compared to full classical execution of the original circuit. 20 and 30 qubit configurations of the random circuit exhibit the case where subcircuit execution is done with classical and quantum resources. in this case there are resource savings in both the maximum number of qubits required as well as the maximum amount of memory required to execute the subcircuits classically. QFT demands full qubit usage as no subcircuits can be classically simulated.
The cost results show that partitioning circuits and utilising hybrid execution can result in efficient use of classical and quantum resources in most cases. These savings are limited by the ability to classically execute subcircuits as these cases demonstrate the benefit of this approach.

\section{Conclusion}
\label{sec:conclusion}

This paper demonstrates that dynamic partitioning and hybrid execution can be beneficial in the partitioning and execution of quantum circuits. Practical use of quantum circuit partitioning will require a dynamic approach to deciding cut points to avoid prohibitive post processing as a result of sub optimal cuts. When used alongside parallel execution, the dynamic partitioning approach efficiently finds near-optimal cut points.
Hybrid execution also proved to be a useful mechanism in reducing noise and distributing work. The result is a cost effective partitioning and scheduling framework that efficiently uses both classical and quantum resources. Given that quantum computation can be expensive, incorporating classical resources into the execution flow allows for easier scalability. Combined with the benefits observed with noiseless classical execution, hybrid execution presents itself as a promising mechanism in executing partitioned quantum circuits.
This paper provides a framework to partition and execute quantum circuits in such a way that it mitigates the effects of error, adapts to circuit structure and executes in a resource efficient manner. With sufficient classical resources, the benefits of classical execution can be extended with contraction path optimisers such as Cotengra along with approximate calculation or tensor slicing techniques. Such resources will be able to handle even more of the workload which will produce better results in executing large quantum circuits.

\end{document}